\documentclass[aps,10pt,prl,twocolumn,showpacs,groupedaddress]{revtex4}  
\usepackage{graphicx}  
\usepackage{dcolumn}   
\usepackage{bm}        
\usepackage{amssymb}   
\usepackage{amsmath}

\begin{document}

\title{On the possibility of a relativistic correction to the E and B fields around a current-carrying wire}
\author{Ron Folman}
	\email{folman@bgu.ac.il}

	\affiliation{Department of Physics, Ben-Gurion University of the Negev, Be'er Sheva 84105, Israel}
\date{\today}

\begin{abstract}
It is well known that electric and magnetic fields may change when they are observed from different frames of reference. For example, the motion of a charged probe particle moving parallel to a current-carrying wire would be described by utilizing different electric or magnetic fields, depending on from which frame of reference the system is observed and described. To describe the situation in all frames by utilizing the theory of relativity, one has to first describe the situation in one particular frame, and this choice in the case of a current-carrying wire is the topic of this paper. Specifically, I consider the question of in which frame the current carrying wire is neutral. The importance of relaxation processes is emphasized. As an example, I examine a specific alternative to the standard choice, and consider its theoretical and experimental validity. An outcome of alternative approaches is that in the rest frame of a wire, running a current introduces also an electric field by giving rise to a minute charge. Present day experimental sensitivities, specifically those of cold ions, may be able to differentiate between the observable signatures predicted by the different approaches.
\end{abstract}

\pacs{37.10.Gh, 32.70.Cs, 05.40.-a, 67.85.-d}
\maketitle

The theory of electromagnetism \cite{maxwell} and the special theory of relativity \cite{einstein} have been confirmed in an abundance of experiments. In the following I describe an effect that is a result of these two theories and which may have been left unnoticed in the experiments performed due to its relatively weak nature.

The interplay between electric and magnetic fields when one moves from one frame of reference to another stands at the base of our understanding of how the theory of relativity affects these fields \cite{jackson}. The special theory of relativity clearly explains how these fields are observed as changing as a function of the frame of observation, but in order to know what the fields are, the fields in one particular frame must first be determined by other theoretical or experimental means. It is the description of this particular frame, usually chosen to be the rest frame, that is at the core of this paper.

The dependence of an elementary charge (e.g. that of an electron or a proton) on its velocity would have been observed in particle accelerators, in the neutrality of atoms or even when heating a metal, and has been investigated by numerous methods (\cite{jackson} and refs. therein). Specifically, experiments intended to measure a second order change of the charge with its velocity, seem to have observed no effect \cite{PRD1977,PLA1991}. It is therefore of common belief that at least in low energies, such a dependence has been ruled out. However, it seems that the dependence of charge density on the mean velocity (so-called drift velocity) in the rest frame of a wire has not been examined thoroughly and therefore cannot be ruled out. In this paper I explain why such a fundamental effect could be considered as possible and calculate the expected signal. (following the above, we shall assume the invariance of the elementary charge with respect to its velocity, meaning that if the total number of particles is conserved then the total charge is conserved.)

More specifically, the E and B fields are a result of the charge density $\rho$ and the current density $j$. It is these physical observables that are analyzed henceforth. As is well known, the 4-vector $(\rho c, {\bf j})$, $c$ being the speed of light, transforms according to the Lorentz transformations when the observer moves from one frame to the other. However, as noted, these transformations may only tell us what the charge and the current are in a specific frame if we have previously determined them in another frame. It is common practice to assume that in the rest frame of a wire (which we define here as the rest frame of the nuclei or the protons that form a wire) a current does not give rise to a charge. This comes about from the assumption that the electrons behave as free particles and therefore accelerating them to some finite velocity does not change the particle-particle distances. If indeed the effect we consider in this paper is real, the latter assumption is not correct.

The question of the charging of a current carrying wire in its rest frame, may be phrased in the following way: Albert Einstein, in his 1905 paper, wrote \cite{LC}: "At v=V, all moving objects - viewed from the "stationary" system - shrink into plane figures" (where V is the speed of light). It is therefore widely believed (although I am not sure if this has been verified) that in accelerators, when clouds of particles with strong enough restoring forces, e.g. an ensemble of nucleons in a nucleus, are accelerated to some finite speed, the cloud length is Lorentz contracted thus producing a pancake shaped cloud in the lab frame of the accelerator \cite{pancake}. In the case of a nucleus, this comes about from the fact that the restoring forces keep its spherical shape in its rest frame. If such a pancake indeed exists, the question can then be asked as to whether such a contraction of an "electron cloud" may occur in a current carrying wire thereby producing a net charge in the wire rest frame.

Following this work \cite{course}, I became aware of numerous previous theoretical presentations of similar ideas and the criticism they received (see Refs. \cite{book,Redzic} and references therein). In Appendix I we give an example of such a theory and the criticism it received. Contrary to some of the previous works (e.g. \cite{tomislav}), the analysis described here will adhere to standard electrodynamics. Also contrary to some previous works which state, to my mind without sufficient proof, that the wire in its rest frame is indeed charged (e.g. \cite{Peters} and more recently \cite{Brill}), it is claimed in this work that not enough is known to arrive at this conclusion. At the same time, it seems there is no theoretical or experimental proof in favor of the standard approach which holds that the wire is neutral in its rest frame.

This paper goes beyond the previous presentations as it emphasizes the importance of internal dynamics, giving rise to restoring (relaxation) processes (as in the pancake example above), and the fact that only a three dimensional many-body covariant analysis of the system can be expected to give a theoretical resolution regarding at which frame the equilibrium is at the $\rho_{net}=0$ point. If this happens in several frames, a situation not compatible with relativity, the question most relevant may be in which frame the restoring dynamics are faster. In this context we analyze in detail Maxwell's equations and Ohm's law. This work also presents a new analysis of the experimental sensitivity required to observe the hypothesized effect and analyzes the feasibility of several novel experimental methods to make such an observation. The analysis indicates that cold ions may serve as a probe which is sensitive enough to differentiate between the different possibilities.

Let us begin directly with pointing out briefly one counter example to the standard approach, following which it is discussed in a detailed and consistent manner. If one assumes (as is commonly done) a zero net charge density for a current-carrying wire in its rest frame $S$ ($\rho_+=|\rho_-|=\rho_0$ where $\rho_0$ is the free electron density per unit length in the material when no current is applied), one must, due to Lorentz transformations, assume a charge density of $(\gamma-1/\gamma) \rho_0$ in frame $S'$ moving at the drift velocity of the electrons in frame $S$ [where $\gamma=(1-v^2/c^2)^{-\frac{1}{2}}$]. Namely, the wire, as observed in $S'$, becomes charged. One intuitive explanation for this apparent charging is that somehow the net charge density is dependent on the velocity of the charges in $S'$ so that $\rho'_+ \ne |\rho'_-|$. This cannot be the right explanation simply because the same velocity difference in $S$ does not give rise to a similar difference of densities in frame $S$. As described below, while the common choice [$\rho_+=|\rho_-|$ in $S$ and $(\gamma-1/\gamma) \rho_0$ in $S'$] is consistent with the special theory of relativity, it may not be the only plausible application of the theory. For example, one could have applied the theory and the arguments in a symmetric way in $S$ and $S'$. As in $S$ the electrons are moving and the positive charges are stationary, one could perhaps hypothesize to have in $S$ a net charge of $\rho_{net}=\rho_0(1-\gamma)$ while in $S'$ one would expect $\rho'_{net}=\rho_0(\gamma-1)$. As will be shown in the following, the latter choice is also compatible with the special theory of relativity. As for small velocities $\gamma=(1+\frac12 v^2/c^2)$, the latter hypothesis gives rise to a second order effect in the electron velocity in the rest frame of the wire.

The difference between the two choices may perhaps be made more apparent by the following "exposition": we simulate a normal conductor with two infinite solid rods, one made of positive ions and one made of negative ions. As we are not interested in absolute charge but only in charge density (per unit length), the boundary conditions are of no importance. While the positive rod is at rest in $S$, a current in $S$ is realized by moving the negative rod with a velocity equal to the electron drift velocity in a normal conductor. If we accept the standard choice for $S$, an observer in $S$ would claim to see no charge. His friend is an observer in $S^*$, a frame in which both rods have the same velocity (in opposite directions). As in $S$ there is a current while there is no charge, Lorentz transformations do not allow for any other frame to observe a charge of zero. This is well known to the observer in $S$ and thus he communicates to his friend that he must be observing a non-zero charge, giving rise to a non-zero electric field. The friend, however, sees a completely symmetrical world. He observes two infinite rods, built exactly the same way with the exception of an opposite charge, going in opposite directions with equal velocity. As he believes the sign of the velocity vector should have no effect on how charge changes with velocity, he concludes that this symmetrical world must give rise to a charge of zero, in contradiction to what his friend communicated to him, and obviously in contradiction to the theory of relativity. Of course, if the non standard choice [i.e. $\rho_{net}=\rho_0(1-\gamma)$] was observed in $S$, Lorentz transformations would have predicted a zero charge density in the symmetrical world observed in $S^*$.

Let us emphasize that the following detailed analysis of the situation in which the $S^*$ frame noted above is the neutral frame, is given here as a mere possible counter example to the standard approach. As in the real life situation of a current carrying wire the $S^*$ frame is clearly not symmetric (masses and interactions are different), it may very well be that another frame eventually presents itself as the unique frame in which $\rho_{net}=0$.

In his famous lectures on physics Feynman explains in detail an example which concerns the electric and magnetic fields produced by a current-carrying wire, as observed from different frames of reference \cite{chapter}. Following Feynman's formalism, I examine the above alternative approach to the problem at hand. If one is not able to determine theoretically a single preferred option, then one will have to resort to experiment.

This paper is structured as follows: First I put forward in simple terms the difference between the two approaches regarding their underlying arguments. I then describe in more detail the standard approach followed by the alternative approach.
I conclude with the theoretical and experimental consequences due to the difference between the two approaches.

\begin{figure}[b]%
\includegraphics[width=\columnwidth]{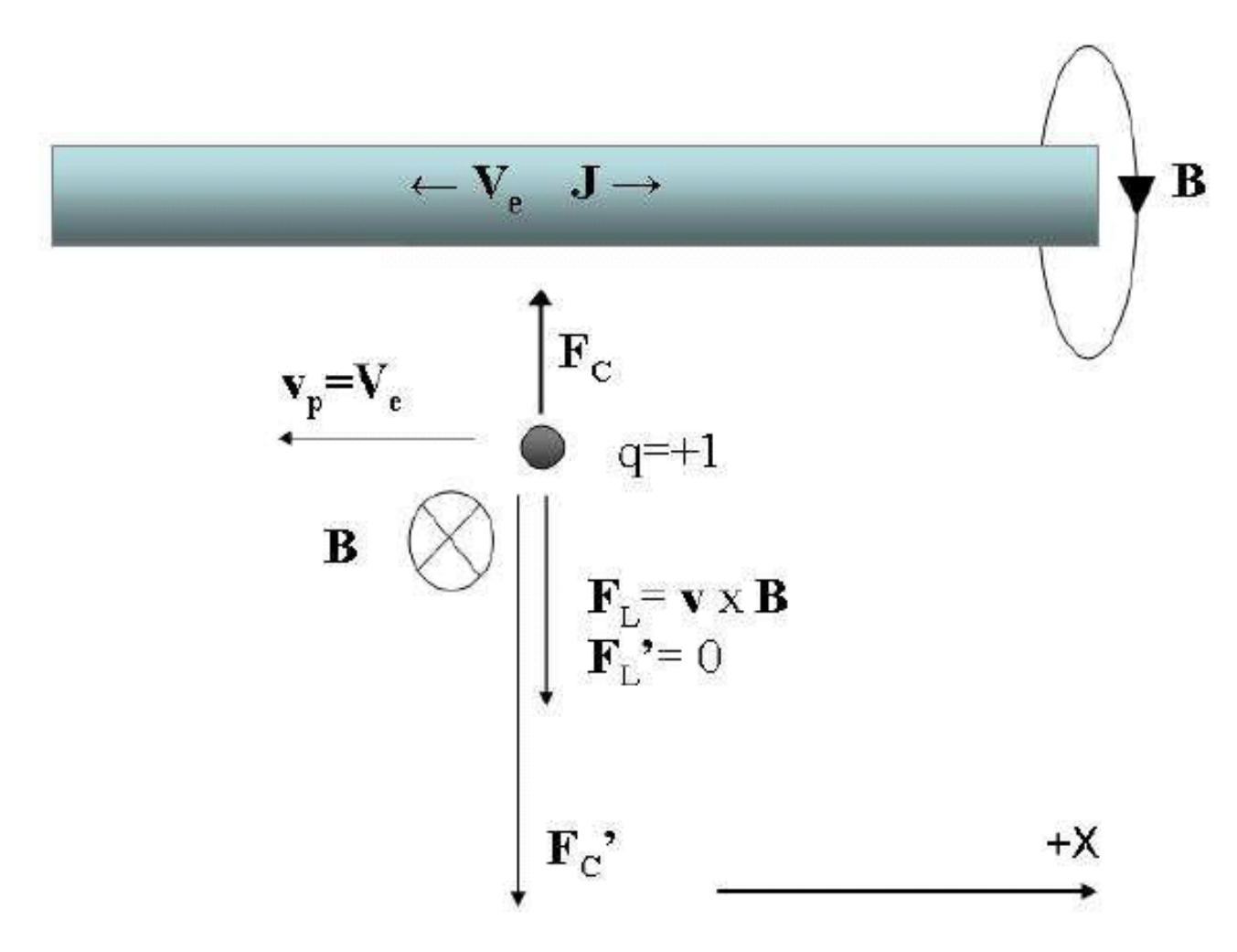}%
\caption{
The system of the wire and probe particle. The wire is stationary in the lab frame S and the particle is stationary in its rest frame S'. The electric ($F_C$) and magnetic ($F_L$) forces in both frames are presented. Note that in S' there is no magnetic force as the particle does not move. Contrary to the standard approach, in the alternative approach presented in this work there is a Coulomb force also in S.
}%
\label{fig:system}%
\end{figure}

Let me first describe the system we are addressing. As presented in Fig. \ref{fig:system}, a current-carrying wire is placed in the lab frame ($S$). Due to the current, the electrons have some mean velocity $v_e$. A positively charged probe particle with charge $q=+1$ moves parallel to the wire axis with velocity $v_p$ and has the same velocity as the electrons so that $v_p=v_e$. It is well known that due to its charge and velocity in the magnetic field of the wire, the particle experiences a magnetic force, the Lorentz force ($F_L$), which is equal in this case to $q v_p B$ and directed away from the wire. The question addressed by Feynman is what would be the forces experienced by the probe particle in its own frame of reference ($S'$), the so-called particle rest frame, namely, in the frame where the probe particle's velocity is zero. While it is clear that the magnitude of the above magnetic force is zero in $S'$, the question is if any other forces are created? Specifically, concerning forces that are perpendicular to $v_p$, one would require, in order to be consistent with the theory of relativity, that the sum of forces in $S'$ ($F'$) would be larger than the sum of forces in $S$ ($F$) by the relativistic Lorentz factor $\gamma$ so that $F'=\gamma F$. This is so because the force is the derivative of the momentum in time, and as we do not expect the momentum in the direction transverse to the relative velocity of the frames (the latter being $v_p$) to change when moving from frame to frame, we expect $F' \Delta t' = F \Delta t$. As the theory of relativity dictates $\Delta t = \gamma \Delta t'$ we therefore require $F'=\gamma F$. Hence, although the magnetic force in $S'$ is zero, the total force in $S'$ cannot be zero.

In his usual brilliant manner, Feynman simply explains that as the wire becomes shorter in frame $S'$, the density of the positive charges in the wire grows and this creates an electric field which gives rise to a Coulomb force ($F_C'$) which acts on the probe particle in $S'$. However, to give rise to an electric field Feynman has to differentiate between the relativistic effect due to the frame change as it acts on the positive and the negative charges. No quantified physical argument is put forward to justify the specific asymmetry of the $(\gamma-1/\gamma)$ factor leading to a non-zero charge density in $S'$. In fact this factor is simply a Lorentz transformation of the assumption of neutrality in the rest frame of the positive charges. As will be explained in the following, the latter assumption is equivalent to the assumption that no restoring forces are at work on the electrons such that the electrons' inter-particle distance in some frame other than $S$ is restored to their original distance in the wire frame before the current was turned on. Put in simple words this means that the electrons behave as free particles. This may be called "The Standard Assumption". As Feynman's treatment following this "assumption" is completely consistent with the theory of relativity, he indeed finds that as required $F'=\gamma F$.

In the alternative approach presented here we assume that both the positive and negative charges have the same restoring forces, so that they maintain their natural inter-particle distances in their rest frame. Following the above "pancake" example, in both frames the density is increased only for the moving particles due to Lorentz contraction. This imposed symmetry may be called "An Alternative Assumption". This approach also leads to $F'=\gamma F$ and hence it is also consistent with the theory of relativity. However, it leads to the charging of the current-carrying wire in its rest frame, a surprising result indeed. Thus, while both approaches are consistent with relativity, they start with very different assumptions and arrive at very different descriptions concerning the physical situation in frame $S$.

It seems likely to this author that both these assumptions present the extreme edges of a wide spectrum of possibilities. What is common to all models occupying this spectrum is that all, but the one in the standard edge, predict the charging of the wire in its rest frame, in varying magnitudes.

Let me now briefly describe the standard approach (in my own abbreviated way). A detailed account may be found in Ref. \cite{chapter}. First, note that the magnetic force $F_L$ experienced by the probe particle in $S$ is

\begin{eqnarray}
	F_L=q v_p B = q v_p \frac{J}{2\pi r c^2 \epsilon_0}= q v_p \frac{\rho_0 v_e}{2\pi r c^2 \epsilon_0}= \nonumber \\
\frac{q \rho_0 v^2}{2\pi r c^2 \epsilon_0},
\label{eq:mag}
\end{eqnarray}
where r is the distance from the wire, $\rho_0$ is the free electron density (per unit length) in the material when no current is applied, and I have taken (following Feynman's example) $v=v_e=v_p$.

Next, Feynman calculates the Coulomb force $F_C'$ in $S'$ by integrating the net charge along the wire axis and its electric field at the position of the particle. Note that Feynman takes the positive charge density to be $\rho_+'=\gamma \rho_0$ and the negative charge density to be $\rho_-'=-\rho_0/\gamma$ so that the net charge density is $(\gamma-1/\gamma) \rho_0=\gamma(1-1/\gamma^2)\rho_0=\gamma\beta^2\rho_0$, where $\beta=v/c$. The Coulomb force comes out to be

\begin{eqnarray}
	F'=F_C'=\gamma\beta^2\rho_0 \int \frac{q}{4\pi\epsilon_0(r^2+x^2)}\frac{r}{\sqrt{(r^2+x^2)}}dx=\nonumber \\
\gamma\beta^2\rho_0 \frac{q}{2\pi\epsilon_0 r} =\gamma F_L=\gamma F,
\label{eq:elec}
\end{eqnarray}
where, according to the standard approach, $F$ and $F'$ are the total forces in $S$ and $S'$, respectively.

Let me now describe in detail the alternative approach. As noted, the assumption regarding the charge neutrality of the wire in $S$ is replaced with a symmetry assumption requiring that $\rho'=-\rho$, where $\rho$ is the net charge density (noted previously as $\rho_{net}$). For simplicity, we shall not write the relevant parameters in units of $\rho_0$ as in the standard approach but rather in units of $\rho_0^{**}=\gamma\rho_0$, where $\gamma=(1-v_e^2/c^2)^{-\frac{1}{2}}$.

To see exactly what the charge density should be in the two frames, let us remind ourselves of the four vector transformations. The vector $(\rho c, j_x)$ transforms under the Lorentz transformation matrix

\[%
\begin{array}
[c]{cc}%
\gamma \left(
\begin{array}
[c]{ll}%
~~1 & -\beta\\
-\beta & ~~1
\end{array}
\right)  .
\end{array}
\]

We therefore see that $\rho' c=\gamma\rho c - \gamma \beta j$. As we are are discussing $\rho$ and $j$ in units of $\rho_0^{**}$ (in which $j$ is simply $v_e$), and as $j$ is in the positive direction while the boost $\beta$ from $S$ to $S'$ is in the negative direction, we find $\rho'=\gamma\rho + \gamma \beta^2$. Imposing symmetry, namely that $\rho'=-\rho$, we find that $\rho=(1/\gamma-1)\rho_0^{**}$ and $\rho'=(1-1/\gamma)\rho_0^{**}$.
Consequently, the net charge in $S'$ is now $(\gamma-1) \rho_0$ contrary to the standard $(\gamma-1/\gamma) \rho_0$. That the positive charge density in $S'$ is $\rho_+'=\gamma \rho_0$ and the negative charge density is $\rho_-'= \rho_0$ is a natural outcome under the assumption of symmetry, as the positive charges are moving and the electrons are at rest. In the same way, in $S$ we now have a net charge of $(1-\gamma) \rho_0$ as the electrons are moving and the positive charges are at rest. Thus, our conclusion concerning the state of the wire in $S$ must be different from that of standard approach that the wire in $S$ is not charged. To make sure that also the alternative approach is consistent with the theory of relativity, let us now calculate the forces and see if we again find $F'=\gamma F$ as required.

In $S$ we now expect to find both a Coulomb force and a magnetic force as depicted in Fig. \ref{fig:system}. As the probe particle is positively charged and as the wire is negatively charged, and as the probe particle is moving in the same direction as the wire electrons, the electric and magnetic forces will be in directions opposite to each other.  Note also that in order to be consistent, the $\rho_0$ appearing in the magnetic force in $S$ must also be boosted by $\gamma$ as it is due to the electrons.
The total force in $S$ will thus be

\begin{eqnarray}
	F=F_L+F_C=\gamma\rho_0 \beta^2\frac{q}{2\pi\epsilon_0 r}+\nonumber \\
(1-\gamma)\rho_0 \int \frac{q}{4\pi\epsilon_0(r^2+x^2)}\frac{r}{\sqrt{(r^2+x^2)}}dx=\nonumber \\
(\gamma \beta^2+1-\gamma)\times\frac{q\rho_0}{2\pi\epsilon_0 r}.
\label{eq:alt1}
\end{eqnarray}

In frame $S'$ we find
\begin{equation}
	F'=F_L'+F_C'=0+(\gamma-1)\times\frac{q\rho_0}{2\pi\epsilon_0 r},
\label{eq:alt2}
\end{equation}
and we therefore find
\begin{equation}
	\frac{F'}{F}=\frac{\gamma-1}{\gamma \beta^2+1-\gamma}=\gamma
\label{eq:comp}
\end{equation}
as required.

Let us now consider the theoretical consequences. One of the most important issues to clarify is how both approaches deal with charge conservation, which we assume to be correct. For example, as the alternative approach claims that in $S$ a current induces a charge density, the conclusion must be that for an infinite wire, an infinite charge must be deposited. This is not very appealing indeed. However, this is also what happens in $S'$ in the standard approach.


As it is not very clear what charge conservation means for an infinitely long wire, and as it is not clear what an experiment verifying any result concerning an infinitely long wire should look like, we shall consider a current in a conducting loop of finite dimensions.

\begin{figure}[b]%
\includegraphics[width=\columnwidth]{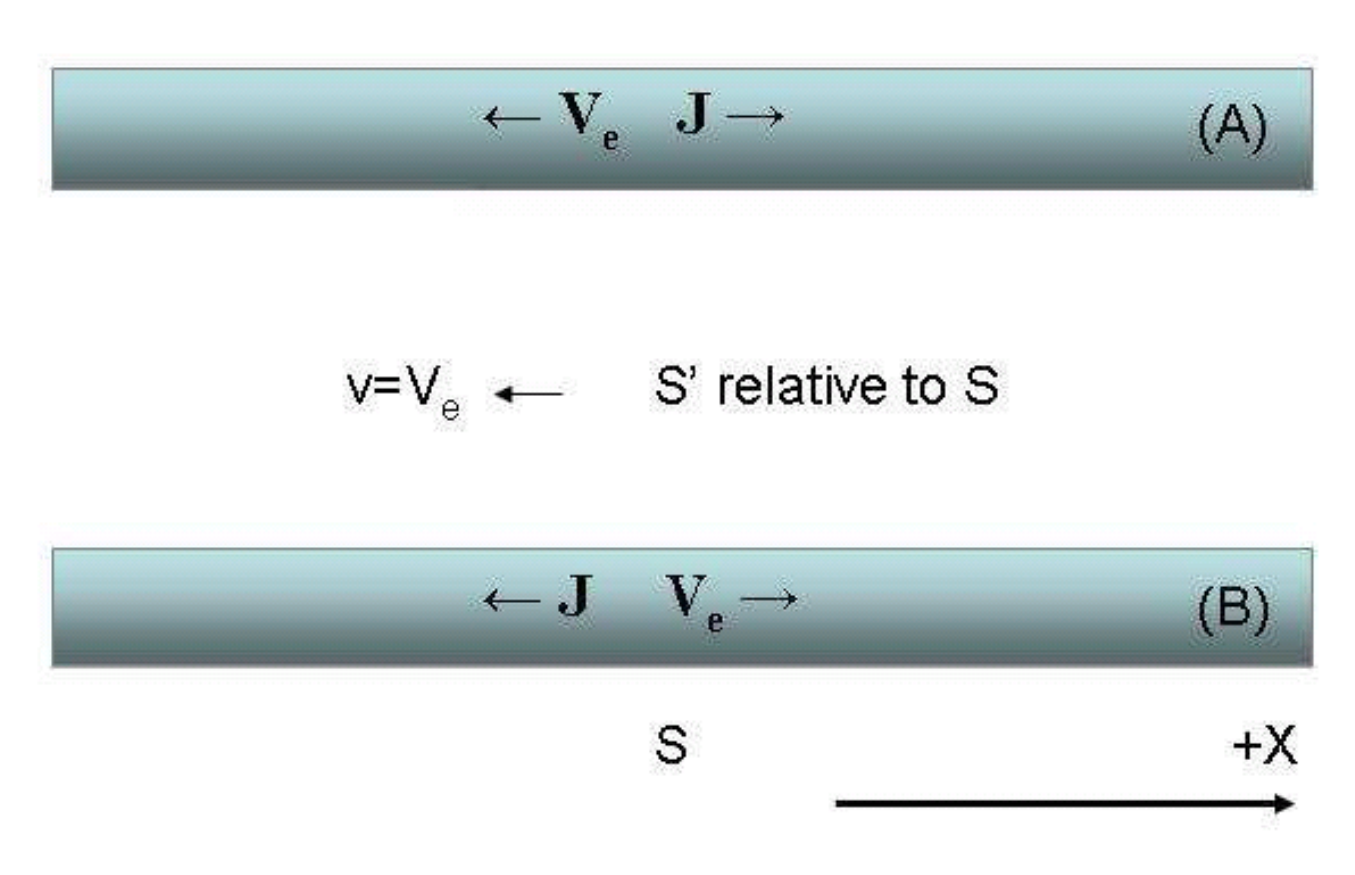}%
\caption{
The current loop composed of two wires (presented) and vertical contacts at their ends (not presented). We also present the two frames of reference: $S$ is the lab frame in which the wire is stationary and $S'$ moves at the velocity of the electrons in wire A.
}%
\label{fig:loop}%
\end{figure}

Let us thus examine a loop in the shape of a square of unit length, so that in essence there are two wires of the form noted above in the $\hat{x}$ direction (see Fig.~\ref{fig:loop}), and between them, at their ends, vertical connections \cite{purcell}. It is clear that the total charge of the two horizontal wires A and B is zero in $S$ for the standard approach and negative, $2(1-\gamma)$, for the alternative approach. We of course expect that in $S'$ both approaches would lead to the same charge the loop had in $S$, as is verified in appendix II. This is of course to be expected as in relativity the number of negative and positive charges is frame independent.

But what about the situation in the alternative approach where the mere fact that we ran a current in $S$ does not adhere to charge conservation? This is perhaps the most far reaching implication of the alternative approach.

In most experimental scenarios a current source (e.g. battery) must be embedded into the loop. An ad-hoc assumption may then be that the current source (acting as a charge reservoir) balances the hypothesized charging by being positively charged. Indeed, the previous analysis of charge conservation in the loop when changing between different frames, indicates, that the balance charge on the current source does not need to change from frame to frame. This is the same in both approaches.

However, in this work, as we aim at a feasible experimental configuration, we will not consider a loop with a current source, as this is known to give rise to a first order effect in the electron velocity \cite{book,standard_inhomogeneity}. The latter charging is a much stronger effect and would mask the signal we are searching for \cite{against first order}.

When there is no current source (e.g. one introduces a current by a varying magnetic flux), and the loop charge must be preserved in $S$ when inducing a current, two processes may take place. First, as by moving from an infinite wire to a finite loop without a charge reservoir we have changed the boundary conditions of the problem, it may be that the restoring forces are now such that the second order effect vanishes in the rest frame. Second, the current density may not be homogeneous throughout the cross-section of the wire such that some parts of the cross-section are negatively charged (where the current runs) and some positively charged, so that integrating over the cross section of the wire retains the charge neutrality. For example, if we examine the known fact that superconducting currents run along the wire edges, the expected increase in charge density where the current runs would mean that the center of the wire becomes positively charged while the edges become negatively charged. One should note that such a scenario of course creates electric fields which aim to counter-act the inhomogeneous charge distribution, but at least in the case of a superconductor, these edge currents are stable. Also in a regular conductor, the well known pinch effect, in which parallel currents attract each other thereby causing the electron density to be larger in the center of the wire, is eventually balanced by electric fields to create a stable non-homogeneous current density. In the experimental section of this paper, we will focus on superconductor loops due to the high electron velocity. High velocities may also be found in other conductors such as Graphene \cite{graphene}.


Let us now discuss the issue of internal dynamics. First, let us briefly remind ourselves of the importance of such internal dynamics in relativity. John Bell, perhaps best remembered for his brilliant work known as "Bell's inequalities", has clearly stated that if two particles (or space ships in his famous paradox \cite{Bell}) are at rest in some system $S$ with a distance $L$ between them, and if in that system they are then accelerated with the same force and for the same time, the distance between them remains a constant. This must be so if relativity is to be consistent with the known laws of motion, namely, $x(t)=x(0)+\int v(t)dt$. So why are physicists, from the heavy ion collision community \cite{pancake} or from other communities (e.g. \cite{mirror}), confident that they should see Lorentz contraction in $S$? The answer is also given by Bell's paradox. If the above two particles retain their distance from one another as $L$, that distance in their new rest frame $S'$ must be $\gamma L$ as relativity demands a factor $\gamma$ between length measurements done in two different frames. So in fact, if these two particles would represent the edges of some system (such as a solid rod) this system would feel stretched in its rest frame, thus giving rise to internal dynamics such as relaxation or restoring forces. If these restoring forces shrink the system back to its equilibrium length $L$ in the new rest frame $S'$, then, the system must also shrink in $S$. A system restored to a sphere in $S'$ (as in the heavy ion collision example) would thus look like a pancake in $S$.

Applying the above insight to the current carrying wire, it seems the solution to the debate would be found if one can decide if the electrons in the ion lattice are closer to a system of isolated (non-interacting) particles, in which case their distance from one another is increased in other frames (e.g. to $\gamma L$ in $S'$), or they are closer to an interacting system with strong restoring forces in some other frame, in which case we would expect their inter-particle distance in $S$ to shrink (e.g. to $L/\gamma$ if their distance in $S'$ is restored to $L$) - once the current is turned on. How then can we try and describe the internal dynamics inside the wire, which we shall name "longitudinal relaxation"? One such description of the internal dynamics is given by Ohm's law, the latter being our next station.

So far, we have considered a symmetric physical system of positively and negatively charged particles (for final remarks regarding the symmetric system see \cite{symmetric}). This is perhaps the situation in a plasma of electrons and positrons (as well as a system of electrons and holes in a semiconductor), or perhaps, as noted in the introduction, in a wire made of two counter propagating solid rods, one made of negative ions and one of positive ions. However, for the case of a normal current carrying wire, one needs to take into account the possible asymmetry caused by the fact that the positive particles are much heavier and they are held together by strong bonds which form a stringent lattice \cite{asymmetry}. One therefore needs to account for internal dynamics in the presence of the above asymmetry. Perhaps the most adequate formalism available for the movement of the electrons within a wire is Ohm's law. Let us therefore analyze Ohm's law in the context of the ideas presented in this paper. Could it be that the simple form of Ohm's law, $\vec{j}=\sigma \vec{E}$, is inconsistent with the $\rho\ne 0$ suggested by the alternative approach?

The covariant form of Ohm's law is \cite{jackson-Ohm}:

\begin{eqnarray}
	j^{\alpha}=\frac{\sigma}{c}F^{\alpha\beta}u_{\beta}+
\frac{1}{c^2}(u_{\beta}j^{\beta})u^{\alpha}
\label{eq:Ohm1}
\end{eqnarray}
where $j^{\alpha}$ is the 4-current, $u^{\alpha}=\gamma(c,\vec{u})$ is the 4-velocity of the medium and $F^{\alpha\beta}$ is the field strength tensor.

This may also be written as (recalling that $j_0=j^0$ and $j_i=-j^i$ for $i=1-3$):

\begin{eqnarray}
	g_{\alpha\beta}j^{\beta}=\sigma F_{\alpha\beta}\beta^{\beta}+\beta_{\alpha}\beta_{\beta}j^{\beta}
\label{eq:Ohm2}
\end{eqnarray}
where $g_{\alpha\beta}$ is the metric tensor and $\beta^{\alpha}=u^{\alpha}/c=\frac{\gamma}{c}(c,\vec{u})$.

In Appendix III we show that this reduces for $\vec{u}=0$ to the known form of $\vec{j}=\sigma \vec{E}$. We also show in the appendix that if $u\ll c$, one finds to first order the current density $\vec{j}=\rho\vec{\beta}+\sigma(\vec{E}+\vec{\beta}\times\vec{B})$, which is the familiar form of the Hall effect.

However, most relevant for our discussion is the charge, namely, $\alpha=0$. One then easily finds that Eq. \ref{eq:Ohm2} reduces to

\begin{eqnarray}
	\rho-\gamma(\gamma\rho-\gamma\vec{u}\cdot\vec{j})=\sigma\gamma\vec{u}\cdot\vec{E}
\label{eq:Ohm3}
\end{eqnarray}

In the rest frame of the wire, where $\vec{u}=0$ and $\gamma=1$, one finds the identity $0=0$ for any charge density $\rho$. This means that Ohm's law puts no constraints on what the charge density of the wire will be in the rest frame of the wire. Indeed, if we run a current through a charged capacitor plate (parallel to its plane) it will obey Ohm's law although the wire is charged. To conclude, it seems as if the alternative approach does not contradict Ohm's law in the simple way examined above, and more elaborate tests of the internal dynamics are required if one is to try and theoretically differentiate between the different approaches.

Let us now also analyze the situation in the direction transverse to the current, namely within the wire cross section, and attempt to analyze the processes dominating the "transverse relaxation". To begin with, one should note that together with the continuity equation Ohm's law (in the wire rest frame) gives
\begin{eqnarray}
{\bf \nabla}(\sigma{\bf E})=({\bf \nabla}\sigma)\cdot{\bf E}+\sigma{\bf \nabla}\cdot{\bf E}=\nonumber\\
{\bf \nabla}\cdot{\bf j}=-\dot\rho,
\label{eq:surface1}
\end{eqnarray}
which together with Gauss's law ${\bf \nabla}\cdot{\bf E}=\rho/\epsilon_0$ gives:
\begin{eqnarray}
({\bf \nabla}\sigma)\cdot{\bf E}+\rho\sigma/\epsilon_0=-\dot\rho,
\label{eq:surface2}
\end{eqnarray}
and as long as $\sigma$ is uniform one finds
\begin{eqnarray}
\dot\rho=-\rho\sigma/\epsilon_0 ~~~~\rightarrow ~~~~~ \rho(t)=\rho(0)e^{-\sigma t/\epsilon_0},
\label{eq:surface3}
\end{eqnarray}
where this calculation does not hold for the surface where ${\bf \nabla}\sigma\ne0$. This seems to indicate that any charge in the bulk of the wire would decay to zero.

The above calculation should be met with some scepticism. First, the $\epsilon_0/\sigma$ time scale of roughly $10^{-20}$ seconds seems to indicate that the above model is not complete. Further doubt should arise when one notes there is no length scale in the equations. Such a time scale means for example that a wire with a radius smaller than the radius of a single atom would already require the electrons to move at a speed faster than that of light! More so, one would expect, for example, that there would be considerable difference in this time scale depending on whether the wire radius is much larger or much smaller than the electron mean free path of a few nano-meters. Furthermore, the idea that the spatial distribution of the extra charge would be infinitely thin (as is implied by the charging of the surface) would give rise to a very high electrostatic energy arising from the electron-electron repulsion. A more complete model would thus give a finite width distribution as a function of the charge. A similar result should come when one takes into account the finite temperature of the electrons from which one would assume that the thermal velocity would cause diffusion of the electrons against any tendency to concentrate them at a specific location.

A more complete description should include the magnetic force $ev\times B$ pushing the electrons towards the center of the wire (which may be termed a self induced Hall effect or a pinch as in plasma physics). One may combine between Eqs. \ref{eq:surface1}-\ref{eq:surface3} and the above, by utilizing ${\bf j}=\sigma ({\bf E} + {\bf v} \times {\bf B})$ and the continuity equation, assuming the conductivity is homogeneous, and noting that
\begin{eqnarray}
-\dot\rho={\bf \nabla}\cdot{\bf j}=\nonumber\\
{\bf \nabla}\cdot\sigma ({\bf E} + {\bf v} \times {\bf B})=\nonumber\\
\rho\sigma/\epsilon_0+\sigma {\bf \nabla}\cdot{\bf v} \times {\bf B}=\nonumber\\
\rho\sigma/\epsilon_0-\sigma{\bf v}\cdot({\bf \nabla}\times{\bf B}).
\label{eq9}
\end{eqnarray}

This gives rise in a steady state ($\dot\rho=0$) to $\rho=\epsilon_0{\bf v}\cdot{\bf \nabla}\times{\bf B}$. The conclusion of Eq. \ref{eq:surface3} is thus not necessarily valid.

In appendix IV we analyze the role of the Lorentz force in different frames. It is shown that if we assume the bulk of the wire in its rest frame has no internal transverse electric field, the Lorentz force in other frames is typically not zero. Hence, if the Lorentz force is considered as the only force acting on the electrons and thus the only force responsible for the steady state, a zero transverse electric field in $S$ does not lead to a transverse steady state in other frames. If in $S$ there is a transverse electric field which leads to a steady state by equaling the above magnetic force, then some form of non zero transverse charge density distribution in $S$ must result. Whether integrating over such a transverse distribution should give an overall neutrality - is another question.

Let us emphasize that Lorentz transformations should exist for every point in the wire cross section. Hence, as the transverse charge distribution changes with time in one frame, it should also change in time in other frames, and as the transverse distribution reaches a steady state in one frame its boosted distribution in another frame (boosting as usual parallel to the current or wire axis) should also present a steady state. Once this steady state is reached, there is no transverse current and ${\bf j}=\sigma({\bf E}+{\bf v}\times{\bf B})=\sigma{\bf E}_\parallel$. It remains to be seen if these transverse processes determine in some way what the allowed charging in the rest frame is or what frame is the neutral frame.


To conclude the theoretical analysis we have made, we may say the following: as a current is made to run in a wire, electrons are accelerated in the rest frame of the wire to their drift velocity. If they are to be considered as an ensemble of free particles, kinematics demands that their particle-particle distances would not change and no extra net charge density would appear. This requires the charging of the wire in other frames. If the electrons in a wire are subjected to significant longitudinal restoring forces, then the wire may be restored to neutrality faster in some other frame, in which case, the wire in its rest frame will have to be charged. Any solution would also need to take into account relaxation forces in the transverse direction. Thus a full three dimensional many-body covariant analysis is required.

Let us now examine the experimental implications with some detail. Let us start with a simple straight copper wire, being part of a square loop circuit, assuming the extra charge is provided by the current source. Copper has a density of $8.94$g/cm³, and an atomic number of $63.546$g/mol, so there are $140685.5$mol/m$^3$. In $1$ mole of copper there are $6.02\times10^{23}$ atoms. Therefore in $1$m$^3$ of copper there are about $8.5\times10^{28}$ atoms. Copper has one free electron per atom, so the electron density is equal to $n=8.5\times10^{28}$ electrons per m$^3$. Let us assume a typical laboratory situation in which we have a current of $I=1$ Ampere in a wire of $1$mm diameter (i.e. a cross section area of $A=7.85\times10^{-7}$m$^2$). The drift velocity can therefore be calculated to be

\begin{eqnarray}
v_e= \frac{I}{nAq} = \frac{1}{(8.5\times10^{28}~7.85\times10^{-7}~1.6\times10^{-19} )}\nonumber \\
= 9.367\times10^{-5} m/s
\label{eq:exp}
\end{eqnarray}

As $\beta=v/c=9.367\times10^{-5}/2.99\times10^8=3.133\times10^{-13}$, we can now calculate the Lorentz factor $\gamma$ to be equal to about

\begin{equation}
\gamma=\frac{1}{\sqrt{1-\beta^2}}\approx 1+\frac{1}{2}\beta^2\approx 1+4.5\times10^{-26}
\label{eq:gamma}
\end{equation}

In a $1$m wire we expect to have $8.5\times10^{28}\times 7.85\times10^{-7}=6.6725 \times 10^{22}$ electrons, so that the excess charge predicted by the alternative approach would be that of approximately $3\times10^{-3}$ electrons per meter or $\rho_{net}=-5\times10^{-22}$C/m. This is perhaps the strongest expected signal, namely that a positively charged probe particle should be attracted to a current-carrying wire, even when the particle is at rest relative to the wire.

It is very hard to answer the question of whether or not previous experiments should have observed such an attraction. Typically, charged particles in the vicinity of a current-carrying wire would be mostly affected by the magnetic Lorentz force. Following Eq. \ref{eq:alt1} and the approximation of $\gamma$ in Eq. \ref{eq:gamma}, one finds that the ratio $R$ between the electric force and the magnetic force is

\begin{equation}
R= \frac{F_C}{F_L}=-\frac{v_e^2}{2c^2}/(\frac{v_e v_p}{c^2}+\frac{v_e^3 v_p}{2c^4}).
\label{eq:ratio}
\end{equation}

This ratio equals approximately half for small velocities and when $v_e=v_p$ and may therefore seem to point to a significant and observable effect. However, using the value for $v_e$ calculated above and taking into account $\frac{1}{2} m_p v_p^2=K_B T$, one finds that even for a heavy probe particle (e.g. an atomic mass of $100$), the particle would have to be at a temperature of $0.1$nK for the above ratio of half to be obtained. At a temperature of $1$mK the effect visibility, namely $R$, is already at the $0.5$\% level. Hence very low velocity charged probe particles are needed and these have been made experimentally available only recently via the trapping of laser cooled ions \cite{cold_ions}. However, these systems do not typically incorporate current-carrying wires close to the charged particles, and in any case, as we show in the following, the absolute effect for an ion of charge $|e|$ next to a normal conductor would be very small.

Such low velocities of probe particles, and indeed where $v_e=v_p$, may be found in a parallel current-carrying wire configuration (e.g. in the setup by which the Ampere standard is determined). However, the overall wire charge is again the very small quantity calculated above and the magnetic Lorentz force acting on all the moving charges in the probe wire would again dominate. Let us calculate the forces involved explicitly. The magnetic Lorentz force between two $1$m wires is $\frac{\mu_0}{2\pi}
\frac{1}{d}I_1 I_2$ which is for $1$A and $1$m distance $2\times10^{-7}$N. On the other hand, the expected electric field E(r) is (Eq. \ref{eq:alt1}):

\begin{equation}
E(r)=(1-\gamma)\times\frac{\rho_0}{2\pi\epsilon_0 r}\approx\frac{\rho_{net}}{2\pi\epsilon_0 r}
\label{eq:elec_field}
\end{equation}
where $\rho_{net}=-\frac{\beta^2}{2}\rho_0$. At $1$m distance, the electric field is about $E=9\times10^{-12}$V/m.

If the second parallel wire has the same current and therefore the same induced charge of $\rho_{net}$, the electric force between the wires will be on the order of $10^{-33}$N, much smaller than the magnetic Lorentz force which the experiment is measuring. One may then wish to simply charge a disconnected piece of wire as a probe to the hypothesized induced charge in the current-carrying wire. Namely, to achieve via a macroscopic system $v_p=0$.

Let us assume this probe charge to be $q_p=1$C and see what forces we may expect. The expected electric force will be $F=E q_p\approx10^{-11}$N. On the other hand, the probe charge will induce a rather strong dipole in the current-carrying wire (as the internal field in the metallic wire, perpendicular to its longitudinal axis, is zero, we neglect the polarization of the atoms). An upper limit on the dipole $d$ per unit length may be estimated as the diameter of the wire ($1$mm) multiplied by the charge of the free electrons per unit length which we have estimated above to be $6.6725 \times 10^{22} ~ 1.6 \times 10^{-19}\approx10^4$C/m. As the force is simply the field gradient times the dipole, we find the force per unit length of the current-carrying wire to be (for a point like probe and without geometrical factors):

\begin{equation}
F=10^4 \times 10^{-3} \times \frac{1}{4\pi \epsilon_0}\frac{d}{dr}\frac{q_p}{r^2}=\frac{10}{2\pi \epsilon_0}\frac{1}{r^3}
\label{eq:diploe_force}
\end{equation}
which is in the closest area to the probe (i.e. distance of $1$m) of the order of $F=1.8\times10^{11}$N, which is obviously huge compared to the induced electric force we would like to measure. In appendix V we show that a lower limit gives $7\times10^9$N.

However, if we choose $q_p=e$, the above lower limit value for the force reduces to about $10^{-28}$N while the predicted electrical force is $q_p E=1.6\times10^{-19}\times 9\times10^{-12}\approx10^{-30}$N, not very much smaller. This may enable a relative measurement although the absolute values are extremely small. As the electric field of the wire scales as $\frac{1}{r}$ while the dipole force scales as $\frac{1}{r^3}-\frac{1}{r^2}$ for the upper and lower limits respectively, reducing the distance will not help.

One may also wish to try and measure the interaction between the electric field induced by the hypothesized charging of the current-carrying wire and the polarization it induces in atoms. In appendix VI, I explain why this is a difficult experiment to perform and consequently why previous experiments may have overlooked the effect. In appendix VII, I explicitly calculate the magnitude of this signal.

The above shows that the effect is quite elusive and it stands to reason that experiments so far have overlooked it. I present in the following an initial feasibility study of one idea that may enable the observation of the effect.

For the sake of this example, let us assume that the distribution of charges in a superconducting wire which is part of a persistent current loop, follows the current density distribution. Namely, that the excess in negative charge is taken from nearby parts of the wire through which current does not run, so that integrating over the cross section of the wire, charge neutrality is maintained. Such a charge distribution may be observable by cold ions. Inhomogeneous charge distributions in superconductors from other effects \cite{SC_inhomogeneity} may also be possible and should of course be analyzed.

As the mean velocity of electrons along the wire axis of a superconductor may be orders of magnitude higher than the typical electron drift velocity in normal conductors (simply because the cross section in which the super current runs is very small), one would expect that due to the alternative approach, there would be an excess of electrons in the edges of the wire. As noted, such an excess would create an excess of positive charges in the center of the wire. The exact current distribution in a superconductor and the calculation of the velocity is not a trivial matter \cite{sc1} but let us make a simplified estimate. A superconductor may carry a current density of $10^6$A/cm$^2$ (or $10^{-2}$A/$\mu$m$^2$), similar to that of a normal wire. Hence, a superconducting strip of $100\mu$m width and $1\mu$m thickness can carry $1$A of current, similar to our example copper wire above. However, due to the concentration of current in the edges, one may roughly estimate that this $1$A of current utilizes a wire cross section of about $1\mu$m$^2$, six orders of magnitude less than in the above copper wire. This determines $v_e$ and $\beta$ to be six orders of magnitude larger than previously. Indeed, the literature discusses velocities as high as $1$km/s \cite{sc2}. Following Eq. \ref{eq:elec_field}, while using the same free electron density as in copper with the reduced cross section, the electric field also becomes six orders of magnitude larger. Hence, at $10\mu$m distance from the wire, one would expect an electric field of about $1$V/m. This field should reverse sign when moving from the edge of the wire to its center. The effect of such a field should be observable if the force of $Eq$ it applies on the ion is similar or at least not much smaller in magnitude relative to the force applied on the ion by the trapping potential. The latter, in the harmonic approximation, is simply $kx$ where $k=m\omega^2$ and $x$ may be estimated as the ground state size $\sqrt{\hbar/m\omega}$. Indeed, for a cold ion trap of $10$MHz frequency, both forces are on the order of $10^{-19}$N. Recently, a force of about $10^{-22}$N induced by an electric field of $1.8$mV/m, was measured by cold ions \cite{cold_ion_force1}. One should note that such a measurement close to a surface may be hindered by so-called "patch potentials" \cite{patch_potentials}. Several methods may be used to bypass this problem; first, cooling to low temperatures has been found to reduce this hindering effect; second, as this effect seems to become weaker with the strong scaling of $1/d^4$, where $d$ is the ion-surface distance, taking the ion-surface distance to $100-1000\,\mu$m would only reduce the signal electric field by one to two orders of magnitude, while reducing the masking fields by considerably more; third, one may oscillate the super current in direction or amplitude to form an oscillating force, and this could help purify the signal \cite{cold_ion_force2}. Finally, the ion should be in a state without a magnetic moment as the force due to the magnetic field induced by the $1$A current, is significant. Noting the above considerations, it seems reasonable that adequate experimental parameter values may be found so as to ensure the visibility of the hypothesized effect.

To summarize, this paper did not attempt to give a final answer, but rather to lay down in a consistent manner (using known electrodynamics and charge conservation) the fundamental aspects of the problem, while introducing several insights beyond previous works concerning relaxation processes and experimental feasibility.

As an example of a possible alternative to the standard approach, I have presented a different approach as to how the physical situation concerning the electric and magnetic fields in the stationary frame $S$ of a current-carrying wire, may be determined. While typically one determines the state of the system to be neutral in $S$ (by assuming electrons are a free ensemble and therefore their particle-particle distances do not change upon acceleration), and makes use of the theory of relativity to calculate the state in other frames, one possible alternative approach begins with a physical symmetry based argument requiring that the restoring forces at work on both positive and negative charges are equal, and utilizes relativistic transformations to arrive at a different description of frame $S$.

In the symmetry based argument leading to neutrality in the "middle man" frame $S^*$, we modeled a current carrying wire as made of two counter propagating solid rods, one made of negative ions and one of positive ions, and where current comes from their relative speeds. It seems there would be a consensus that the system is completely symmetric and the charge density should be zero. Hence, if it is indeed found that the total charge density is zero in $S$ or some frame other than $S^*$ (in which the two rods have the same velocity), the theoretical analysis should be able to explain quantitatively what the difference is between the latter rod model and the real-life system of a current carrying wire in its rest frame and in the frame where it is neutral.

The main message of this work is three fold: First, it seems that both theoretical and experimental statements made so far, have not been able to sufficiently prove in favor of one resolution or another. Second, it seems that the rules of relativity alone may not be sufficient to solve the debate. A theoretical attempt to decide which approach is favorable, should take into account internal restoring (relaxation) processes, both transverse and longitudinal, in order to find out in which frame the equilibrium is at the $\rho_{net}=0$ point. If this happens in several frames, a situation not compatible with the known transformations between frames, the question most relevant may be in which frame the restoring dynamics are faster. Only a full three dimensional many-body covariant analysis could perhaps decisively bring a theoretical resolution. The answer may very well be system and preparation dependent and will thus have to take into account specific boundary conditions. The third message is that an experimental feasibility study indicates that cold ions may serve as a probe which is sensitive enough to differentiate between different possibilities.

It may eventually be found that theory does not favor a specific approach \cite{Redzic's conclusion}, and choosing between them experimentally will perhaps be a formidable, yet as we have shown - possible, task.

\section{Acknowledgements}

I am sincerely thankful to Daniel Rohrlich, Yonathan Japha, Ferdinand Schmidt-Kaler, Michael Gedalin and Valery Dikovsky for their critical review of this work, and for helpful discussions. Special thanks to Ashok K. Singal, who was kind enough to repeat his arguments after nearly 20 years. Finally, I deeply thank my dear friend and colleague Carsten Henkel with whom I have had many enlightening discussions. This work was supported by the German-Israeli fund (GIF) as part of a project dedicated to cold ions near charge and current carrying structures.\\

{\bf Appendix I: A proposal from the 1990s}

Following this work \cite{course}, I became aware of a previous presentation of a similar idea two decades ago \cite{tomislav} and the criticism it received \cite{ashok}. In this paper we therefore re-visit the idea and introduce several novelties: first, while the previous work was criticized as being "unpalatable" \cite{ashok1} as it used theoretical and textual language beyond what is common practice, this work analyzes the problem within the established and accepted language and laws of physics (e.g. conservation laws). For example, the previous work states that standard electrodynamics is incorrect \cite{tomislav1}, while this work does not support such a claim. Furthermore, the significant difference between the previous formulation of the problem as well as the proposed solution, and the present one, results in different experimental predictions.

Let us note in some detail five differences between the present and the previous work:

\begin{itemize}

\item	In \cite{tomislav1} it is claimed that the standard electrodynamics theory is incorrect: "We have proved that in SET (standard electrodynamics) the law of invariance of charge is not well founded…". The present paper makes no such claim. In this paper I claim SET is just as founded in both of the competing alternatives.  Namely, I claim that there are several Lorentz invariant options to choose from.
\item	In \cite{tomislav1} new and perhaps debatable theoretical constructions such as "Lorentz invariant charges" are defined, whereas in this paper only the standard elements of electrodynamics are used, and it is shown that the total charge of a current loop is conserved between frames in both theories. In the criticism made by \cite{ashok1} it is said: "We find Ivezic's modified definitions of the charge invariance and of the charge neutrality themselves to be unpalatable". Indeed while the previous proposal aims at charge neutrality for a wire in all frames, the present proposal makes no such claim.
\item	In \cite{tomislav2} it is claimed that there is no electric force between two parallel current carrying wires in the rest frame. In this paper it is claimed that this force is not zero. Because Ref. \cite{tomislav2} utilizes both Lorentz contraction of the wire and Lorentz enhancement of the electron density at the same time, the two effects cancel each other and the neutral wire cannot feel the electric field of the other wire. I believe that in \cite{tomislav1,tomislav2} there is confusion between the role of the two Lorentz effects. In the criticism made by Ref. \cite{ashok1} it is said: "But in order to measure the total charge in a section, he simultaneously employs two different stretches of the conductor length in any given reference frame, one for the ions and another for the electron subsystem". The same confusion is apparent in Eq. (3) of Ref. \cite{tomislav2}, where the magnetic force between two parallel wires is calculated.
\item	Due to the same confusion, when in Ref. \cite{tomislav3} the situation in a stationary superconducting loop is addressed, it is said that: "This causes the moving-electron subsystem to shrink to a smaller length in the laboratory frame...the electric field caused by N moving electrons situated on a smaller, contracted, ring...". On the contrary, in this paper, no Lorentz contraction of the ring in which the electrons move is hypothesized; only the contraction of the mean distance between electrons. What is assumed is that as the current is concentrated at the edges of the superconductor, these edges will become negative while the center will become positively charged. As in the previous item, this is another experimental prediction that is different from that presented in  \cite{tomislav1,tomislav2,tomislav3}.
\item	Contrary to the previous work, this paper does not require that the basic law of charge conservation be broken, as it does not claim that a bare loop becomes charged when current is made to flow.
\end{itemize}

It is therefore quite obvious that the formulation of the problem and the proposed solution presented in this paper are very different than those presented in the early 1990s.

In addition, this paper adds several new considerations. First, the previous work did not consider the asymmetry of the system due to the fact that the protons or positively charged centers that form the wire's solid lattice are much heavier than the conducting electrons. This work considers this through an analysis of Ohm's law. Most importantly, this work emphasizes the importance of longitudinal and transverse relaxation forces. Second, this work analyzes the feasibility of novel experimental techniques in observing the hypothesized effect.\\

{\bf Appendix II: Charge conservation}

Let us describe what happens in $S'$ in the standard approach. The vector $(\rho_{net} c, j_x)$ transforms under the Lorentz transformation matrix we have presented earlier so that in the standard approach, as $\rho_{net}=0$ in $S$ for any wire, one finds $\rho'_{net}=-\beta j_x \gamma /c$. As $j_x$ for wires A and B in $S$ is equal and opposite, it is clear that in any other frame the net charges on the two wires would be the same but with opposite sign, and so the total charge of zero is conserved in the standard approach. For completeness I also arrive at this conclusion by a direct calculation, presented below, where, as expected, the calculation reveals that the standard choice leads in $S'$ to an equal but opposite charge density in the two wires, so that the total charge in the complete loop is maintained at zero.

Similarly, the alternative approach also leads in $S'$ to the same charge observed in $S$. This is also presented below. The total charge density on both wires comes out to be $2(\gamma-\gamma^2)$ and taking into account Lorentz contraction, the total charge on the two wires is $2(1-\gamma)$ which is equal to their charge in $S$. The vertical connections remain unchanged.

For a direct calculation, let us first remind ourselves why $\rho'=\rho\gamma_0'/\gamma_0$ where $\rho$ and $\rho'$ are the apparent charge densities, and where $\gamma_0$ and $\gamma_0'$ are the Lorentz factors due to the particles' velocity relative to the two frames of reference. Let us take the Lorentz transformation of the electric field stating $E_\perp'=\gamma^* (E_\perp+ v \times B_\perp)$ \cite{jackson} (in the case where $\beta$ is perpendicular to $E$), where $\gamma^*$ is due to the velocity $v=\beta c$ between the frames, and apply it to two oppositely charged parallel plates lying in the xz plane and moving with velocity $v_x=v_0=\beta_0 c$. Let us also note that all Lorentz factors are due to movement in the x direction. The above transformation is now $E_y'=\gamma^*(E_y-vB_z)$ and as $B_z=\rho v_0 /c^2 \epsilon_0$, the transformation takes the form $E_y'=\rho'/\epsilon_0=\gamma^*\rho/\epsilon_0(1-vv_0/c^2)$. Hence $\rho'/\rho=\gamma^*(1-vv_0/c^2)$. As $\gamma_0'=\gamma_0 \gamma^* (1-\beta\beta_0)$ one finds that $\rho'/\rho=\gamma_0'/\gamma_0$ or $\rho'=\rho\gamma_0'/\gamma_0$.

In the standard approach, the charge density of the first wire (A) in $S'$ is proportional to $(\gamma-1/\gamma)$. The positive charges in the second wire (B) on the other side of the loop have exactly the same velocity relative to $S'$ as those in wire A and so their charge density is proportional to $\gamma$ just as it is in wire A. In $S'$, the electrons in wire B have a velocity of $2v/(1+\beta^2)$ relative to the electrons in wire A, and so one may describe the situation of the electrons in wire B as having $\gamma_0=1$ in their rest frame and $\gamma_0'=(1+\beta^2)/(1-\beta^2)$ in $S'$. As  $\rho=-1/\gamma$, their charge density in $S'$ should be proportional to $-\gamma_0'/\gamma$. Hence the charge density in wire B on the other side of our loop is simply proportional to $\gamma-\gamma_0'/\gamma=1/\gamma-\gamma$ which is just the negative of the charge density in wire A. Consequently, the total charge on the two wires in the standard approach is zero also in $S'$.

In the alternative approach the charge density of wire A in $S'$ is $\gamma-1$ (in units of $\rho_0$). Again, also the positive charges on wire B would experience a $\gamma$ factor just as it is in wire A. In $S'$, the electrons in wire B have again a velocity of $2v/(1+\beta^2)$ relative to the electrons in wire A, and so again $\gamma_0'=(1+\beta^2)/(1-\beta^2)$ and their charge density should thus be proportional to $-\gamma_0'$. Hence the charge density in wire B is simply $\gamma-\gamma_0'$. Adding the charge density of the two wires one finds that the total charge density is $2(\gamma-\gamma^2)$, and dividing by the Lorentz contraction one finds that the total charge is the same as it was in $S$.\\

{\bf Appendix III: Covariant Ohm's law}

In this appendix we briefly examine the covariant form of Ohm's law.

Starting with,

\begin{eqnarray}
	g_{\alpha\beta}j^{\beta}=\sigma F_{\alpha\beta}\beta^{\beta}+\beta_{\alpha}\beta_{\beta}j^{\beta}
\label{eq:Ohm2_app1}
\end{eqnarray}
where $g_{\alpha\beta}$ is the metric tensor, $j^{\beta}$ is the 4-current, $\beta^{\alpha}=u^{\alpha}/c=\frac{\gamma}{c}(c,\vec{u})$ is the 4-velocity of the medium and $F^{\alpha\beta}$ is the field strength tensor.

Taking $\alpha\ne0$ one finds
\begin{eqnarray} \vec{j}-\gamma\vec{\beta}(\gamma\rho-\gamma\vec{\beta}\cdot\vec{j})=\sigma\gamma(\vec{E}+\vec{\beta}\times\vec{B})
\label{eq:Ohm2_app2}
\end{eqnarray}

If $\vec{u}=0$ one finds $\vec{j}=\sigma\vec{E}$. If $|\vec{u}|\ll c$, and one takes the first order in $\gamma\approx 1+\frac{1}{2}\beta^2$, one finds $\vec{j_{tot}}=\vec{j_e}+\vec{j_m}=\rho\vec{\beta}+\sigma(\vec{E}+\vec{\beta}\times\vec{B})$, where $j_e$ and $j_m$ are the dominant terms in the so-called Galilean electric ($E\gg cB$) and magnetic ($cB\gg E$) limits.
\\

{\bf Appendix IV: The Lorentz force in different frames}

Here, as an example, I calculate the Lorentz force on an electron inside a current carrying wire, in the direction perpendicular to the current, if there is no transverse electric field in the wire rest frame. The boost between frames is in the direction of the current. If we expect, in other frames, to reach a steady state in the transverse charge distribution, the transverse Lorentz force $F_{\perp}$, must be zero (if we assume it is the only force acting on the electrons).

Let us denote the wire axis as $\hat{x}$ so that the current density is $j \hat{x}$ (the total current is I). The wire has cylindrical symmetry and its radius is $R$. For simplicity (and hopefully without loss of generality) we look at an electron situated on the $\hat{y}$ axis having a distance $0< r <R$ from the center of the wire, so that the probe point is $r \hat{y}$ ($z=0$). This electron is moving with the current at a drift velocity. We will work in cgs units.

We first note that in the wire rest frame the electric fields at the probe point are (we denote $x,y,z$ as $1,2,3$):
\begin{eqnarray}
E_1=Const\nonumber\\
E_2=0\nonumber\\
E_3=0
\label{eq1}
\end{eqnarray}
where $E_1$ drives the current and $E_{2,3}=0$ because there is no charge inside the wire and due to Gauss's law.

Similarly, the magnetic fields are:
\begin{eqnarray}
B_1=0\nonumber\\
B_2=0\nonumber\\
B_3=\frac{\mu_0}{2\pi}Ir/R^2
\label{eq2}
\end{eqnarray}
where from Ampere's law $\int \vec{B}\cdot\vec{dl}=\mu_0I_c$ with $I_c/I=\pi r^2/\pi R^2$. As $\vec{I}=I\hat{x}$, the velocity of our probe electron is $\vec{v_e}=v_e(-\hat{x})$ and we find that the Lorentz force $ev_e\times B$ is {\it away} from the surface and towards the center of the wire i.e. in the direction of $-\hat{y}$. This is to be expected as what we have calculated is analogous to the attraction between two parallel currents. It is however an interesting result in the fact that the Lorentz force works against the charge tendency to concentrate at the edges.

Let us now find the electric and magnetic fields in other frames. We denote $V$ (or $\beta=V/c$) as the boost velocity. Lorentz transformations of the fields are:
\begin{eqnarray}
E'_1=E_1=Const\nonumber\\
E'_2=\gamma(E_2-\beta B_3)=-\gamma\beta \frac{\mu_0}{2\pi}Ir/R^2\nonumber\\
E'_3=\gamma(E_3+\beta B_2)=0
\label{eq3}
\end{eqnarray}
and
\begin{eqnarray}
B'_1=B_1=0\nonumber\\
B'_2=\gamma(B_2+\beta E_3)=0\nonumber\\
B'_3=\gamma(B_3-\beta E_2)=\gamma\frac{\mu_0}{2\pi}Ir/R^2.
\label{eq4}
\end{eqnarray}

As the total Lorentz force is $F=q(E+\frac{1}{c}v\times B)$ we have
\begin{eqnarray}
F_{\perp}=e(E'_2+\frac{1}{c}v'_e\times B'_3)=-\frac{e\mu_0}{2\pi}\frac{Ir}{R^2}\gamma(\frac{v'_e}{c}+\beta),
\label{eq5}
\end{eqnarray}
where $v'_e$ is the electron velocity in the new frame.

According to relativity $v'_e=(v_e-V) / (1-v_eV/c^2)$ and so for the Lorentz force to be zero we need to demand
\begin{eqnarray}
-\frac{1}{c}\frac{v_e-V}{1-v_eV/c^2}=\frac{V}{c}\nonumber\\
OR\nonumber\\
v_e=0.
\label{eq6}
\end{eqnarray}

This outcome means that if we have current and we want a steady state in all frames, we must have transverse electric fields in the wire rest frame.

What happens when our assumption $E_2=0$ is not made? From symmetry arguments one may conclude that there is a radial electric field all around the cross-section of the wire. Then, if we introduce a closed sphere of surface $A$ into the bulk of the wire and recall that $E_1$ is homogeneous, we may use the integral form of Gauss's law $\epsilon_0\int {\bf E} \cdot {\bf dA}=\rho$ to conclude that the region internal to the sphere is charged.

Assuming $E_\perp\ne0$, the conclusion of a non uniform charge density can be made quantitative if a steady state exists in the rest frame, namely (up to factors of $c$), $E_\perp=-{\bf v}\times {\bf B}$. Since ${\bf \nabla}\cdot{\bf E_\parallel}=0$ and ${\bf v}$ is constant, we may write:
\begin{eqnarray}
\rho=\epsilon_0 {\bf \nabla}\cdot{\bf E_\perp}=\epsilon_0 {\bf v}\cdot ({\bf \nabla}\times{\bf B}).
\label{eq7}
\end{eqnarray}
where the last equality comes from the fact that $\rho=\epsilon_0 {\bf \nabla}\cdot{\bf E_\perp}= - \epsilon_0 {\bf\nabla}\cdot{\bf v}\times{\bf B}=
-\epsilon_0{\bf B}\cdot {\bf \nabla}\times{\bf v} + \epsilon_0 {\bf v}\cdot
{\bf \nabla}\times{\bf B}$ \cite{Rosser}. ${\bf \nabla}\times{\bf v}=0$ is termed an "irrotational flow" (no turbulence) and may also come from the assumption that $v$ is proportional to $E$ ($j=\sigma E$) and that ${\bf \nabla}\times{\bf E}$ is the time derivative of the magnetic field which is zero in steady state.

As Ampere's law implies ${\bf \nabla}\times{\bf B}=\mu_0 {\bf J}=\mu_0 \rho_- {\bf v}$, where $\rho_-$ is the charge density of the free electrons in the steady state, we have:
\begin{eqnarray}
\rho=\rho_+ + \rho_-=\epsilon_0 \mu_0 \rho_- v^2=\rho_- v^2/c^2,
\label{eq8}
\end{eqnarray}
namely $\rho_-=-\rho_+/(1-v^2/c^2)=-\gamma^2\rho_+$, which means that there is charging, or at least a non-uniform transverse charge density.
\\

{\bf Appendix V: A lower limit on the induced dipole of a wire}

We have made a rough estimate for an upper limit of the dipole of a metallic wire. This estimate is perhaps too large as it takes the maximal possible value independent of the size of the inducing charge. Let us now derive a lower limit by making use of the known induced dipole in a grounded metallic sphere. We simulate our $1$mm diameter wire by a line of spheres with radius $a=0.5$mm, with their centers along the x axis at locations $x_i$. We place the charge on the z axis at $z_0$. According to Eq. 2.6 in Jackson \cite{jackson2} the force between a probe charge $q_p$ and a single sphere is expected to be:

\begin{equation}
|F|=\frac{q_p}{4\pi\epsilon_0}\frac{a q_p}{r_i}\frac{1}{[r^2_i(1-\frac{a^2}{r^2_i})^2]}
\label{eq:one_sphere}
\end{equation}
where $r_i=\sqrt{x^2_i+z^2_0}$ is the distance between the probe charge and the center of the sphere, the second ratio is the image charge, and the last ratio is simply the Coulomb force factor of one over distance square, where the distance is calculated between the probe charge and the image charge in the sphere.

If we omit terms of $a/r_i$ with a high power, as we assume $a\ll z_0$, and we also position spheres symmetrically along the x axis so that only the vertical force component counts, we find the total force to be:

\begin{equation}
F_{tot}\approx\Sigma_i\frac{q^2_p}{a^2 4\pi\epsilon_0}\frac{a^3}{r^3_i}\frac{z_0}{r_i}=\Sigma_i\frac{q^2_p}{4\pi\epsilon_0}\frac{a z_0}{r^4_i}\approx \int^{\infty}_{-\infty} \frac{dx}{2a}\frac{q^2_p}{4\pi\epsilon_0}\frac{a z_0}{r^4_i}.
\label{eq:many_sphere}
\end{equation}

We then find that

\begin{eqnarray}
F_{tot}\approx \frac{q^2_p z_0}{8\pi\epsilon_0}\int^{\infty}_{-\infty}\frac{dx}{(x^2+z^2_0)^2}=\frac{q^2_p}
{8\pi\epsilon_0 z^2_0}\int^{\infty}_{-\infty}\frac{du}{(1+u^2)^2}=\nonumber\\
=\frac{q^2_p}{16\epsilon_0 z^2_0}.~~~~~~~~~~~~~~~~~~~~~~~~~~~~~~~~~~~~~~
\label{eq:final_sphere}
\end{eqnarray}

Putting in a probe charge of $1$C at a distance of $1$m, we find a force of $7\times10^9$N, only about two orders of magnitude smaller than our upper limit value. Note that in this approximation the force is not dependent on the diameter of the wire.\\

{\bf Appendix VI: The experimental challenge of induced polarization in the electric field of a wire}

Atom chips are devices in which isolated ultra cold atoms (in vacuum) are typically trapped in magnetic traps, microns away from the surface of the chip. Current-carrying wires on the surface of the chip provide the trapping magnetic fields. The magnetic fields interact with the magnetic moment of the neutral atoms (see our review \cite{review}).

In a recent experiment we have carried out with an atom chip, we have also induced an electric field which interacts with the induced polarizability of the atoms, so that the neutral atoms interact with both fields simultaneously \cite{PRL2003}.

In general, aside from simple situations such as between capacitor plates, electric fields are quite hard to accurately engineer as they very much depend on where the ground is or in general, what are the potential surfaces in the vicinity of the charge. In our atom chip work where we combined electric and magnetic fields to create a lattice of traps, we have concentrated thousands of electrons on our surface electrodes in the vicinity of the neutral atoms. The charging and relatively high electric field were achieved by utilizing a capacitor configuration on the surface of the chip. The typical parameters of the setup we have used were $100-500$V on a capacitance of about $10^{-17}$F coming from an electrode area of $2\times100\mu$m$^2$ and an electrode distance of a few tens of microns. The electric force is attractive and attempts to pull the atoms towards the surface while the magnetic force creates a barrier against such a crash. The magnetic force is simply $F_m=\mu_B \times dB/dr$, where $\mu_B=1.4\times 10^6$Hz/G is the Bohr magneton and the typical magnetic gradient at an atom-surface distance of $50-100\mu$m is $1-10$kG/cm. This gives a magnetic force of about $10^{-22}$N. As the electric force was seen to change the magnetic potentials but on the other hand was not able to crash the atoms to the chip surface, it is clear that its magnitude was only about $10^{-22}$N or less.

As noted, the charging was made possible due to a capacitor configuration. The capacitor also created quite a strong field of about $10^6$V/m, which gave rise to a potential modulation of about $100\mu$K in the atom magnetic trapping potential. Having in a similar configuration (e.g. a current-carrying wire close to a ground potential) an excess of $10^{-3}$ electrons, as expected by the alternative approach, would give rise to a field which is smaller by $6-7$ orders of magnitude and consequently to a potential modulation which is smaller by $12-14$ orders. This would not be observable.\\

{\bf Appendix VII: Possible experiments involving induced polarization}

Let us now explicitly calculate the electric force acting on a neutral atom next to a current-carrying wire (in SI units). We use the previous calculations of the hypothesized field induced by a wire carrying $1$A of current. The field we found in Eq. \ref{eq:elec_field} was $E(r)=9\times10^{-12}\frac{1}{r}$V/m. As the electric potential is just $V(r)=-\frac{1}{2}\alpha E^2(r)$, where $\alpha$ is the atomic polarizability which is typically $5\times10^{-39}$Cm$^2$/V and $\alpha E$ is the induced dipole, we expect a force of:

\begin{equation}
F=-\frac{1}{2}\alpha\frac{dE^2(r)}{dr}=81\times10^{-24}~2.5\times10^{-39}\frac{1}{r^3}
\label{eq:pol_force1}
\end{equation}
which, at an atom-distance of $1\mu$m, should amount to about $2\times10^{-43}$N, an extraordinarily small number. The potential modulation (in degrees Kelvin), mentioned in appendix VI and equal to $\Delta T=-\frac{1}{2}\alpha E^2(r)/k_B$, where $k_B=1.38\times10^{-23}$J/K is the Boltzman factor, would also be much too small to be detectable, even with ultra cold atoms.

Let us make the same calculation for the induced dipole of a neutral metallic wire of unit length. In Eq. \ref{eq:diploe_force} we have found that the induced dipole $d$ of such a wire is bounded from above by $d=1$Cm per unit length. For the purposes of this example, we use this value as our estimate for the dipole strength. Using the same E(r) noted above we find the force on a piece of wire of unit length to be:

\begin{equation}
F=d\frac{dE(r)}{dr}=9\times10^{-12}\frac{1}{r^2}
\label{eq:pol_force2}
\end{equation}
which, at a distance of say $1$m (the Ampere standard setup), amounts to about $10^{-11}$N, four orders of magnitude smaller than the force the Ampere standard setup was meant to measure. However, moving the wires to a distance of $1$mm (again, current is running in only one wire), the force would now have a magnitude of $10^{-5}$N, and this may be measurable. However, while the upper limit used here scales as $q$, the lower limit (as calculated in appendix V) scales as $q^2$, thus giving rise for the expected $10^{-3}$ electrons to a difference of $10^{-22}$ in the magnitude of the force.

\bibliographystyle{apsrev4-1} 

\begin{thebibliography}{10}
\bibitem{against first order} If one does want to use a current source, there may be several way to differentiate between the first and second order signals. For example, charges due to the first order effect are expected to vary longitudinally and furthermore they are expected to change sign with current direction and hence average to zero in an AC current. On the contrary, the charges hypothesized by the second order effect should not vary longitudinally and should not change sign with current direction. In addition, if indeed first order surface charges are required to maintain a current or an energy flow, they are not expected to decay with time in any frame in which there is a current. On the contrary, the total charging per unit length hypothesized here should decay, at least in some frame, where the relaxation processes push the wire towards neutrality. These differences may enable to observe the much smaller relativistic effect.
\bibitem{symmetric} Imagine if you will the replacement of a normal metallic wire by two solid rods, one made of positive ions and the other of negative ions. A current carrying wire may then be simulated by simply moving the negative rod with the drift velocity $v_e$ in the frame $S$ of the positive rod. Now let us construct a frame $S^*$ moving at velocity $V$ relative to $S$, in which both rods move in opposite directions with the same velocity. Due to Lorentz transformations, this velocity will not exactly be $v_e/2$ but that is of no importance. (utilizing the rule for velocity transformations $u'=(u-V)/(1-uV/c^2)$ the reader may easily convince himself that $V$ is simply the solution of the quadratic equation $v_e/c^2 \times V^2 + 2\times V + v_e=0$). As the situation in $S^*$ is completely symmetric, it does not matter if a velocity affects charge density or not; it is clear that what ever the physics is, the total charge density is zero. The only assumption required is that the sign of the velocity vector does not change the charge density in magnitude or sign. Utilizing the rule which has been presented in appendix II, namely $\rho'=\rho\gamma_0'/\gamma_0$, where $\rho$ and $\rho'$ are the apparent charge densities, and where $\gamma_0$ and $\gamma_0'$ are the Lorentz factors due to the particles' velocity relative to the two frames of reference, one easily finds that transforming to $S$, the charge density there is proportional to $1-\gamma$ as predicted by the alternative approach.
\bibitem{asymmetry} Another type of asymmetry which should perhaps be considered lies in the fact that only for the solid lattice of the positive charges may it be claimed that there exists a frame in which all the particles in the closed loop are at rest. No such frame exists for the electrons, when the current is non-zero. However, this is exactly the situation in a ring accelerator and nevertheless we expect to see Lorentz contraction of a nucleus everywhere along its circumference. In this sense, the loop wire may be viewed as a ring accelerator. It is therefore unclear why this fact should favor the standard approach over the alternative approach.
\bibitem{Redzic's conclusion} The conclusion of Redzic's paper \cite{Redzic} that "it appears that the problem of in what frame is an infinite current-carrying wire neutral does not have a unique solution" emphasizes the importance of the experimental considerations detailed in this work.
\end{thebibliography}

%

\end{document}